\documentclass[pdflatex, sn-nature]{sn-jnl}

\usepackage{graphicx}%
\usepackage{multirow}%
\usepackage{amsmath,amssymb,amsfonts}%
\usepackage{amsthm}%
\usepackage{mathrsfs}%
\usepackage[title]{appendix}%
\usepackage{xcolor}%
\usepackage{textcomp}%
\usepackage{manyfoot}%
\usepackage{booktabs}%
\usepackage{algorithm}%
\usepackage{algorithmicx}%
\usepackage{algpseudocode}%
\usepackage{listings}%
\usepackage{float}
\usepackage{caption}
\usepackage{array}
\usepackage{tabularx}
\usepackage{adjustbox}
\usepackage{booktabs}
\usepackage{import}
\usepackage[
disable
]{endfloat}
\usepackage{fancyhdr}

\DeclareCaptionLabelFormat{adja-page}{\hrulefill\\ \textbf{#1 #2} \emph{(previous page)}}

\newcommand{\beginsupplement}{%
        \setcounter{table}{0}
        \renewcommand{\thetable}{S\arabic{table}}%
        \setcounter{figure}{0}
        \renewcommand{\thefigure}{S\arabic{figure}}%
     }
\newcolumntype{N}{>{\centering\arraybackslash}m{.5in}}
\newcolumntype{G}{>{\centering\arraybackslash}m{2in}}

\begin{document}
\title[CIBRA: Computational Identification of Biologically Relevant Alterations]{CIBRA identifies genomic alterations with a system-wide impact on tumor biology}

\author[1,2,3]{\fnm{Soufyan} \sur{Lakbir}}

\author[1,2]{\fnm{Caterina} \sur{Buranelli}}

\author[2]{\fnm{Gerrit A.} \sur{Meijer}}

\author[1]{\fnm{Jaap} \sur{Heringa}}

\author*[2]{\fnm{Remond J. A.} \sur{Fijneman}}\email{r.fijneman@nki.nl}

\author*[1,3]{\fnm{Sanne} \sur{Abeln}}\email{s.abeln@uu.nl}

\affil[1]{\footnotesize \orgdiv{Bioinformatics group}, \orgdiv{Department of Computer Science}, \orgname{Vrije Universiteit Amsterdam}, \orgaddress{\street{De Boelelaan 1105}, \city{Amsterdam}, \postcode{1081HV}, \state{Noord-Holland}, \country{The Netherlands}}}

\affil[2]{\footnotesize \orgdiv{Translational Gastrointestinal Oncology group}, \orgdiv{Department of Pathology}, \orgname{Netherlands Cancer Institute}, \orgaddress{\street{Plesmanlaan 121}, \city{Amsterdam}, \postcode{1066CX}, \state{Noord-Holland}, \country{The Netherlands}}}

\affil[3]{\footnotesize \orgdiv{AI Technology for Life group}, \orgdiv{Department of Information and Computing Sciences, Department of Biology}, \orgname{Utrecht University}, \orgaddress{\street{Heidelberglaan 8}, \city{Utrecht}, \postcode{3584CS}, \state{Utrecht}, \country{The Netherlands}}}


\abstract{
\textbf{Background}: Genomic instability is a hallmark of cancer, leading to many somatic alterations. Identifying which alterations have a system-wide impact is a challenging task. Nevertheless, this is an essential first step for prioritizing potential biomarkers. We developed CIBRA (Computational Identification of Biologically Relevant Alterations), a method that determines the system-wide impact of genomic alterations on tumor biology by integrating two distinct omics data types: one indicating genomic alterations (e.g., genomics), and another defining a system-wide expression response (e.g., transcriptomics). CIBRA was evaluated with genome-wide screens in 33 cancer types using primary and metastatic cancer data from the Cancer Genome Atlas and Hartwig Medical Foundation. \textbf{Results}: We demonstrate the capability of CIBRA by successfully confirming the impact of point mutations in experimentally validated oncogenes and tumor suppressor genes. Surprisingly, many genes affected by structural variants were identified to have a strong system-wide impact (30.3\%), suggesting that their role in cancer development has thus far been largely underreported. Additionally, CIBRA can identify impact with only ten cases and controls, providing a novel way to prioritize genomic alterations with a prominent role in cancer biology. \textbf{Conclusions}: Our findings demonstrate that CIBRA can identify cancer drivers by combining genomics and transcriptomics data. Moreover, our work shows an unexpected substantial system-wide impact of structural variants in cancer. Hence, CIBRA has the potential to preselect and refine current definitions of genomic alterations to derive more nuanced biomarkers for diagnostics, disease progression, and treatment response.
CIBRA is available at \url{https://github.com/AIT4LIFE-UU/CIBRA}
}


\maketitle

\clearpage\section{Introduction}\label{introduction}

Cancer is characterized by genomic instability leading to many somatic alterations, ranging from single nucleotide variants (SNVs) to large-scale somatic copy number aberrations (SCNAs) and structural variants (SVs) \cite{hanahan_hallmarks_2011, martinez-jimenez_compendium_2020}. While the majority of alterations have no defined impact on tumor biology, a few alterations contribute to the development and progression of cancer \cite{martinez-jimenez_compendium_2020}. Computationally identifying these key somatic alterations with a major impact on tumor biology is challenging \cite{martinez-jimenez_compendium_2020, dressler_comparative_2022, porta-pardo_comparison_2017, ostroverkhova_cancer_2023}. Generally, there are two types of computational approaches for identifying biologically relevant alterations: frequency-based methods and impact prediction methods. 

Frequency-based methods such as MutSigCV \cite{lawrence_mutational_2013}, OncodriveFM \cite{gonzalez-perez_functional_2012}, OncodriveCLUST \cite{arnedo-pac_oncodriveclustl_2019}, and MutSig-CL \cite{lawrence_discovery_2014} identify biologically relevant alterations through enrichment of alterations, for example, at the population level or sublocalized within a protein \cite{porta-pardo_comparison_2017, vogelstein_cancer_2013, dees_music_2012, niu_protein-structure-guided_2016}. They rely on the rationale that tumorigenesis follows a Darwinian evolution characterized by variation and selection \cite{martinez-jimenez_compendium_2020, porta-pardo_comparison_2017}. 
However, frequency-based methods rely on large cohorts to identify biologically relevant alterations. This leaves the potential role of many low-frequency alterations in cancer to be discovered. Replication timing, chromatin structure, methylation status, or low-complexity regions can all influence the rate of alterations throughout the genome, leading to hot spots, such as fragile sites for SVs. Due to these factors, low-frequency alterations and SVs pose a challenge for frequency-based methods \cite{sherman_genome-wide_2022, ostroverkhova_cancer_2023}. 

Impact prediction methods such as SIFT \cite{ng_sift_2003}, PolyPhen \cite{adzhubei_method_2010}, or the more recent AlphaMissense \cite{cheng_accurate_2023} can effectively estimate the potential impact of variants on the protein and, its pathogenicity \cite{ng_sift_2003, adzhubei_method_2010, cheng_accurate_2023}.  However, these methods can only assess the impact of missense SNVs within a coding region, making it impossible to assess the impact of, for example, SVs and non-coding variants.

In this work, we explore the idea of using gene expression levels to systematically assess the biological impact of genomic alterations; this approach can address many of the shortcomings listed above. The transcriptome is a phenotypic representation of the cellular system, which can reflect the status of cellular processes, as well as tissue composition \cite{guinney_modeling_2014, way_machine_2018, li_identification_2021, knijnenburg_genomic_2018, kang_prediction_2020}. The key idea is to observe if a genomic alteration is associated with a system-wide change in gene expression levels. Previously, we have shown that transcriptomics data can be used to predict system-wide changes, such as genomic instability, by training a random-forest model to predict the tumor break load \cite{lakbir_tumor_2022}. Moreover, the work from Crawford \textit{et al.} (2022) and Aben \textit{et al.} (2018) highlighted that among omics data modalities, the transcriptome can most closely reflect the system-wide change caused by genomic alterations \cite{crawford_widespread_2022, aben_itop_2018}. As such, we hypothesize that biologically relevant genomic alterations elicit characteristic changes through the system, reflecting the genomic change, whereas alterations without an impact will have no defined systemic change on the system. By assessing the degree of change in the system, we can determine the extent of impact of a genomic alteration. This approach can complement current methods in identifying the impact of low-frequency alterations and SVs, which can aid our understanding of cancer biology, consequences for clinical behaviour, and ultimately personalised care \cite{van_belzen_structural_2021, doig_findings_2022}. 

In this study, we introduce a novel computational method, CIBRA (Computational Identification of Biologically Relevant Alterations), that identifies the system-wide impact of genomic alterations by integrating genomics with transcriptomics data. First, to validate if CIBRA can identify known tumor suppressor genes and oncogenes, we performed a pan-cancer genome-wide screen of coding SNVs in primary cancers \cite{grossman_toward_2016}. Next, to assess the impact of SVs in cancer, we conducted a genome-wide screen on genes affected by SVs in metastatic colorectal, breast, and lung cancer \cite{priestley_pan-cancer_2019}.  In addition, we explored the capability of CIBRA to identify the most impactful type of alteration and gene subregion of known oncogenes and tumor suppressor genes. Lastly, we show an additional utility of CIBRA, the similarity score, that assesses the categorical similarity in terms of the observed system-wide impact between different variants within the same gene, or between different genes.

\section{Methods}\label{sec11}

\subsection{CIBRA scores}\label{section_cibra_score}
\begin{figure*}[!t]
\centering
\captionsetup{font = footnotesize}
\includegraphics[width=\textwidth]{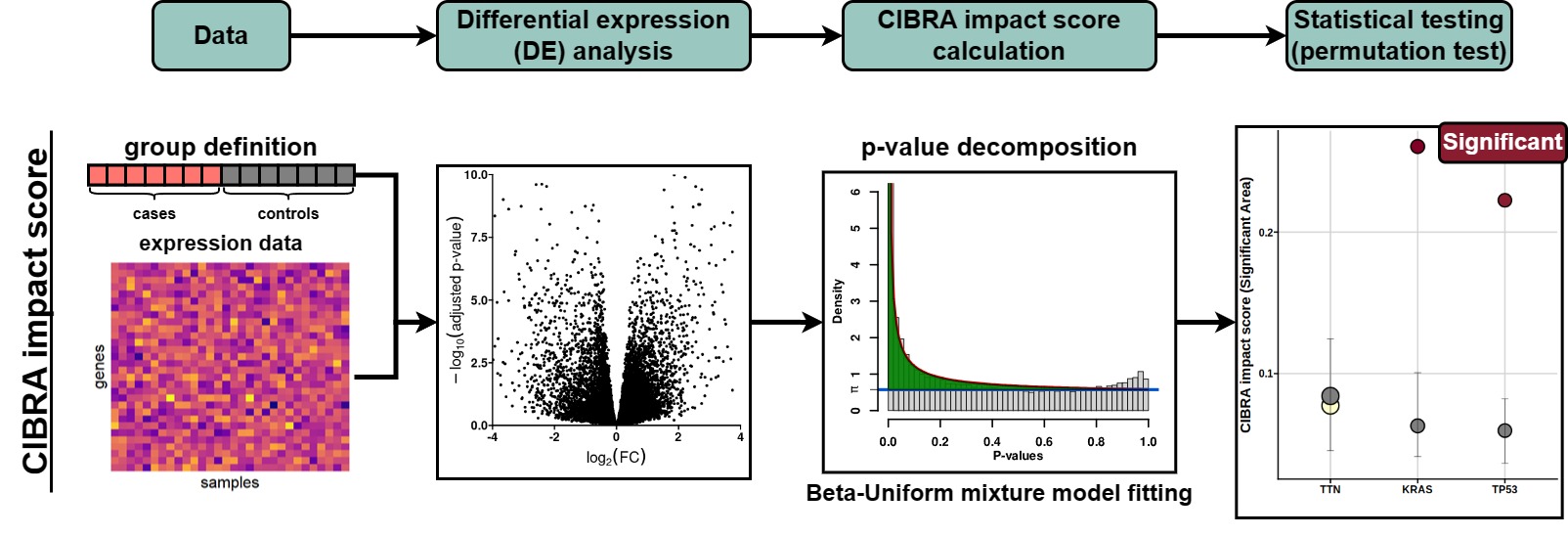}
\caption{Flowchart of the CIBRA impact score calculation. As input CIBRA takes expression data and a group definition in terms of cases and controls: for example, mutated vs. WT samples for a specific genomic alteration. These inputs are used to perform differential expression (DE) analysis between the cases and controls. The p-value distribution generated from the DE analysis is subsequently decomposed by fitting a Beta-Uniform mixture model. From the model, the CIBRA impact score termed the significant area is calculated by taking the integral between the Beta and Uniform components of the mixture model, depicted as the green area in the third panel. Finally, the statistical significance of the impact measure is assessed with a permutation test.}
\label{fig:full_flowchart}
\end{figure*}

\begin{figure*}
\centering
\captionsetup{font = footnotesize}
\includegraphics[width=\textwidth]{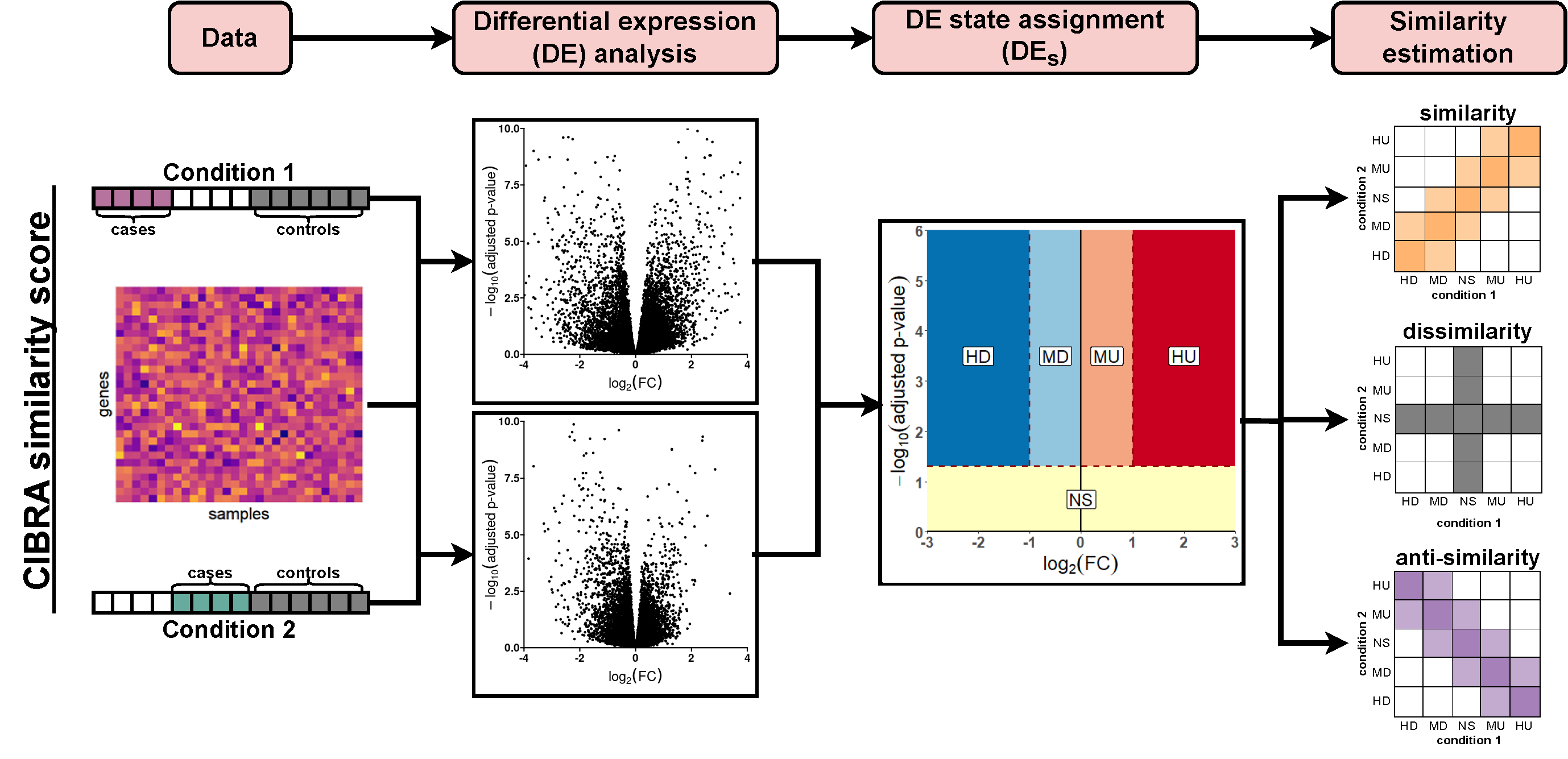}
\caption{Flowchart of the CIBRA similarity score calculation. Gene expression data and group definitions with shared controls were used to perform differential expression (DE) analysis. Using the generated $log_2$ fold changes (FC) and adjusted p-values, five DE states ($DE_s$): highly upregulated (HU), moderately upregulated (MU), moderately downregulated (MD), highly downregulated (HD) and not significant (NS) were assigned per condition for each gene given the corresponding p-value and fold change. With the vectors of $DE_s$, a similarity and anti-similarity score is calculated between the conditions and visualized using a similarity matrix. A permutation test is performed to estimate the significance of the similarity and anti-similarity scores.}
\label{fig:full_flowchart_bottom}
\end{figure*}

CIBRA probes system-wide responses based on samples with (cases) and without (controls) genomic alterations using a Beta-Uniform mixture model \cite{pounds_estimating_2003} to decompose the p-value distribution generated from differential expression (DE) analysis. CIBRA has two scores, the CIBRA impact score (Fig. \ref{fig:full_flowchart}) and the similarity score (Fig. \ref{fig:full_flowchart_bottom}). For the CIBRA impact score, two measures are calculated: the significant area between the Beta and Uniform component of the Beta-Uniform mixture model and the proportion of p-values below a given p-value threshold $\tau$. The statistical significance of the impact measures is assessed by performing 1000 sample permutations and assessing the observed impact score with respect to the permutation distribution (Fig. \ref{fig:full_flowchart}). For the CIBRA similarity score, group definitions with shared controls are used to perform differential expression analysis. The generated $log_2$ fold changes and adjusted p-values are used to define differential expression (DE) states. These states assign genes in whether they are significantly up- or down-regulated or if they are not significantly changed. From the list of DE states, a similarity and anti-similarity score is calculated between the conditions (Fig. \ref{fig:full_flowchart_bottom}). The significance of the scores is assessed with a permutation test.

\subsubsection{Differential expression analysis}
To assess the system-wide response of biological alterations, first, a differential expression (DE) analysis was performed. Any DE analysis method that outputs a valid p-value distribution for omics data, i.e. a p-value distribution that is uniform under the null hypothesis can be used with CIBRA. In this work, DESeq2 (version 1.38.2, default parameters, \cite{love_moderated_2014}), edgeR (version 3.40.1, default parameters, \cite{robinson_edger_2010}) and Limma-Voom (version 3.54.0, default parameters) were assessed. The results reported within this work have been generated using DESeq2, as we have observed that it tends to mostly give valid p-value distributions and is capable of being executed in parallel. EdgeR tended to also mostly show a valid p-value distribution. Limma-Voom gave more invalid p-value distributions with an inflation of 1 using our data compared to the other two methods.  Zero variance and low count genes ($<$ 10 total counts) were excluded from the analysis. 

\subsubsection{Beta-Uniform mixture model}
To estimate the system-wide effect size of a genomic alteration, the p-value distribution derived from DE analysis was decomposed using a Beta-Uniform mixture model \cite{pounds_estimating_2003}. The model is a composite of the Beta distribution with parameter $\beta=1$ and a uniform distribution. The mixture model has a probability density function (pdf), which can be calculated as: 

\begin{equation}
f(x \mid \alpha,\lambda) = \lambda + (1 - \lambda) \alpha x^{( \alpha - 1 )} \quad for\; x,\lambda,\alpha\; \in\; (0,1) \label{eq1}
\end{equation}

where the probability of x is dependent on the shape parameters $\lambda$ and $\alpha$ for $x, \lambda, \alpha \in (0,1)$. The Beta-Uniform mixture model was fitted using the R package BioNet (version 1.58.0, \cite{beisser_bionet_2010}).  

For the Beta-Uniform mixture model, under the null hypothesis that the genomic alteration has no system-wide expression change, the model will lead to a Uniform function with $\lambda = 1$. Under the alternative hypothesis that the genomic alteration has a system-wide expression change, the function will acquire density for small p-values and $0 \leq \lambda < 1$. The larger the signal, the larger the density the function will acquire. Given these hypotheses, we have derived two signal measures that capture the system-wide expression change: the CIBRA impact score termed the \emph{significant area} and the \emph{proportion}.

\textbf{CIBRA impact score (significant area):} The significant area is the integral between the Beta and Uniform components calculated as: 
\begin{equation}
    \int_{0}^{1}f(x)dx = \alpha(1-\lambda)(x^{\alpha-1}-1) \quad for\; \lambda,\alpha\;\in\;(0,1) \label{eq2}
\end{equation}
where $\lambda$ and $\alpha$ are the estimated shape parameters of the Beta-Uniform mixture model. The significant area indicates the extent of p-values that arise from the alternative component, indicating the extent of change in the system. Under the null hypothesis of no system-wide change, the significant area is expected to be 0, while it increases with an increasing change in the system up to a theoretical maximum of 1 (Fig. \ref{fig:full_flowchart}A).

\textbf{CIBRA impact score (proportion):} the proportion of p-values smaller than a significant threshold value ($\tau$). In this study, the threshold $\tau$ was set to 0.1.  The proportion has been taken as a signal measure to accommodate and detect biases in the p-value distribution. If the p-value distribution shows a shift in values toward 1, given the characteristics of p-values, the distribution is deemed invalid. The proportion reflects this bias, as a shift in p-values toward 1 results in fewer p-values below $\tau$, because if there is no signal in the data, a p-value distribution tends to behave uniform. As such, the proportion should be $\tau$ under the null hypothesis, and if below $\tau$, is an indication of an invalid p-value distribution. The proportion can range from 0 to 1.

The significant area is the measure that describes the extent of the system-wide impact and will be referred to as the CIBRA impact score in this manuscript, while the proportion can give an indication of the extent of significant changes in the system. In addition, the proportion can be used to determine whether the observed values are valid.

\subsubsection{Permutation test}
\begin{figure*}
    \captionsetup{font = footnotesize}
    \centering
    \includegraphics[width=0.7\textwidth]{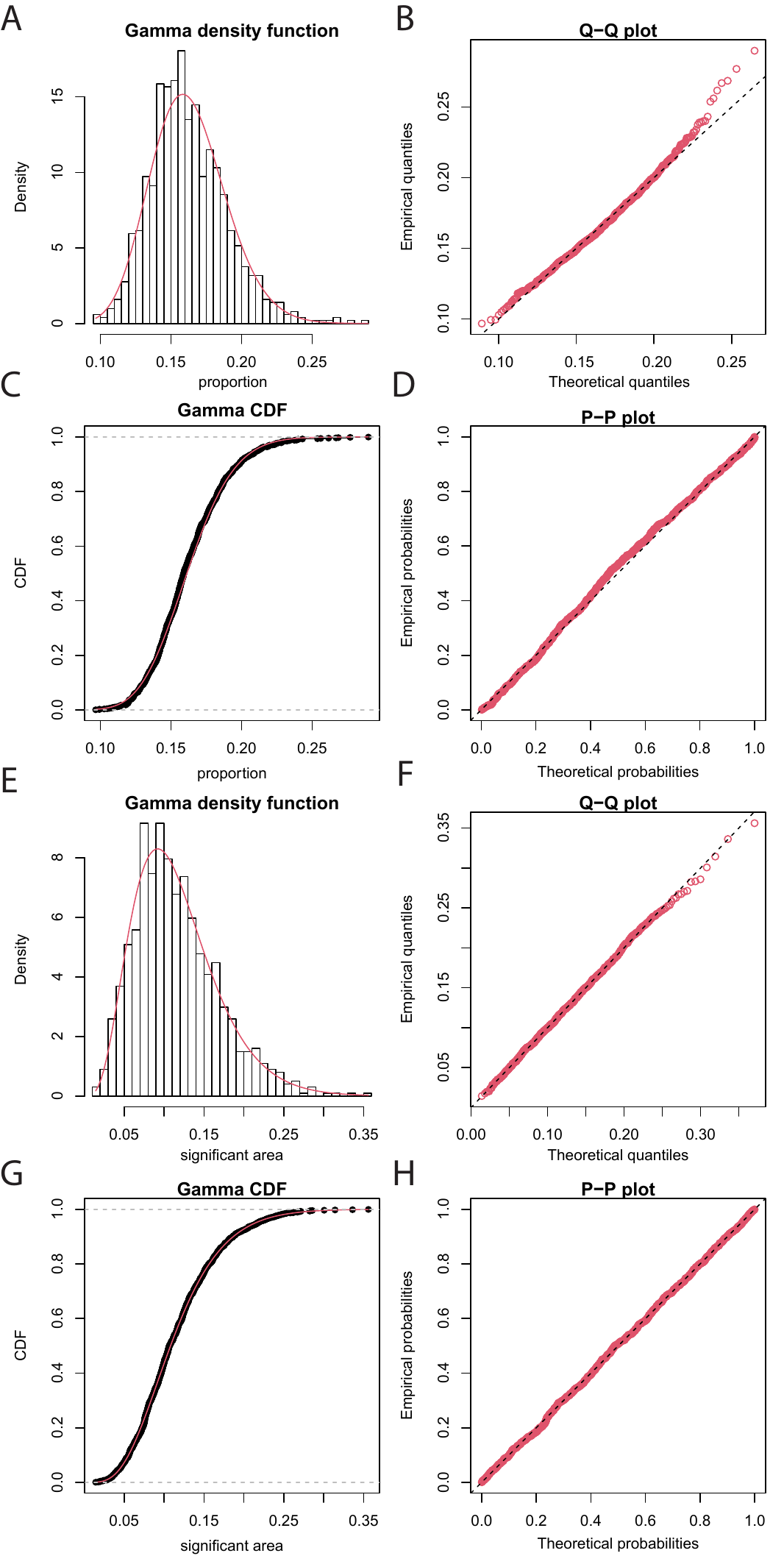}
    \caption{Distribution characteristics of the significant area and proportion from 1000 sample permuted data fitted by a Gamma distribution. \textbf{A)} Histogram of the sample permutation distribution of the proportion with a Gamma density function overlaid. \textbf{B)} Q-Q plot showing the theoretical quantiles of the fitted Gamma distribution plotted against the empirical quantiles of the proportion from the sample permuted data. \textbf{C)} The fitted Gamma cumulative density function (CDF) plotted against the proportion. \textbf{D)} P-P plot of the theoretical probabilities of the fitted Gamma distribution against the Empirical probabilities of the proportion. \textbf{E)} Histogram of the sample permutation distribution of the significant area with a Gamma density function overlaid. \textbf{F)} Q-Q plot showing the theoretical quantiles of the fitted Gamma distribution plotted against the empirical quantiles of the significant area from the sample permuted data. \textbf{G)} The fitted Gamma cumulative density function (CDF) plotted against the significant area. \textbf{H)} P-P plot of the theoretical probabilities of the fitted Gamma distribution against the Empirical probabilities of the significant area. The distribution characteristics show that a Gamma distribution fits well with the CIBRA impact scores: significant area and proportion.}
    \label{fig:sub_2}
\end{figure*}

\begin{figure*}
    \centering
    \captionsetup{font = footnotesize}
    \includegraphics[width=1\textwidth]{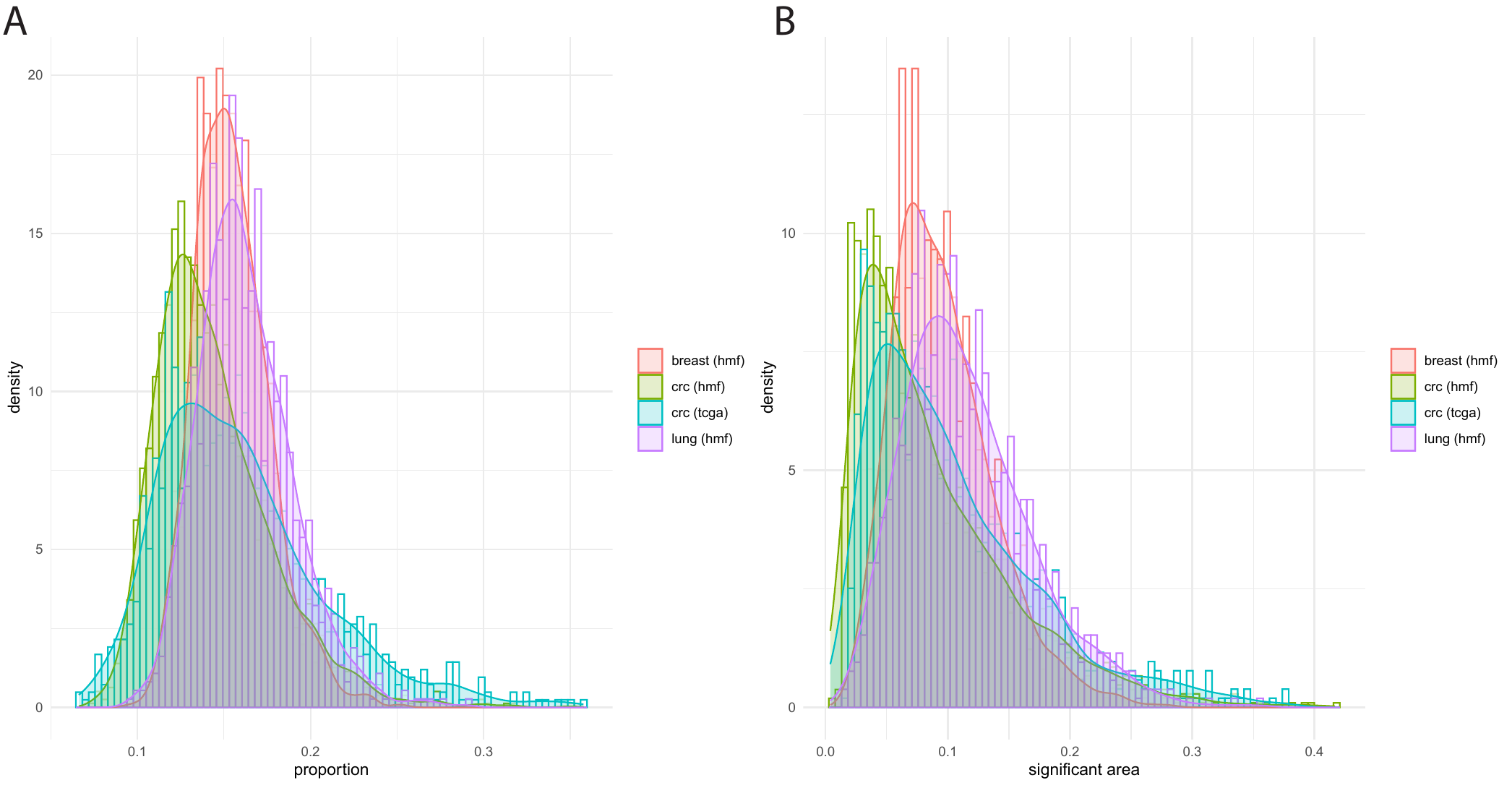}
    \caption{Permutation distribution of the signal measures for different cancer types and datasets. \textbf{A)}Histogram overlaid with the density of the distribution for the proportion calculated for 1000 sample permutations for primary (crc\_tcga) and metastatic colorectal (crc\_hmf), breast (brca\_hmf), and lung (luad\_hmf) cancer. \textbf{B)} Histogram overlayed with the density of the distribution for the significant area calculated for 1000 sample permutations for primary (crc\_tcga) and metastatic colorectal (crc\_hmf), breast (brca\_hmf), and lung (luad\_hmf) cancer. Cancer type has an influence on the shape and position of the permutation distribution. The more heterogeneous the cohort, the wider the distribution. This is especially clear for the primary colorectal cancer dataset consisting of multiple stages of the disease.  As such, generating the sample permutation distribution for each cancer type and dataset is advised.}
    \label{fig:sub_3}
\end{figure*}

To estimate the significance of the CIBRA impact scores given the variation present in the dataset, a permutation test was performed. For all data platforms (TCGA and HMF) and cancer types, at least 1000 sample permutations were made with a case-control parameter grid with 30 steps for the case size and 5 steps for the control size, both starting at 10 and increasing up to the maximum number of samples. The overall distributions for the CIBRA impact scores were found to be well-fitted by a gamma distribution, as shown in Fig. \ref{fig:sub_2}. The cancer type and data platform are confounding factors shifting the distribution (Fig. \ref{fig:sub_3}; p $<$ 0.0001) when taken in a Gamma regression model fitted using a generalized linear model (GLM) with family Gamma and an inverse link function. As such, the permutation distribution must be generated for each cohort (i.e., for each cancer type). The influence of the number of cases and controls on the characteristics of the permutation distribution was also assessed. Given that the case and control sizes are general properties shared between cancer types and data platforms, the influence of case and control sizes has been assessed on one data set. A GLM with family Gamma and an inverse link function was fitted on the CIBRA impact scores calculated from metastatic colorectal cancer data (HMF). No significant relationship was found between the number of cases and controls and the CIBRA impact scores. As such, a generic permutation distribution could be made for each cohort. However, a low number of cases and controls does result in more invalid CIBRA impact scores, i.e., a proportion below $\tau$, meaning an invalid p-value distribution, as shown in Fig. \ref{fig:sub_4}. Moreover, for less than 8 cases/controls, DESeq2 failed to perform DE analysis on our data. As such, we recommend at least 10 cases/control to have reliable results.
To estimate the significance of the signal measures, the signal measures are compared to a Gamma distribution fitted against the corresponding 1000 permutations of the cancer type and data platform. The fit was estimated using the R package fitdistrplus (version 1.1-8, \cite{delignette-muller_fitdistrplus_2015}, parameters: distr="gamma").

\subsection{CIBRA similarity score}\label{section_similarity}
\begin{figure*}
\centering
\captionsetup{font = footnotesize}
\noindent\makebox[\textwidth]{\includegraphics[width=1.3\textwidth]{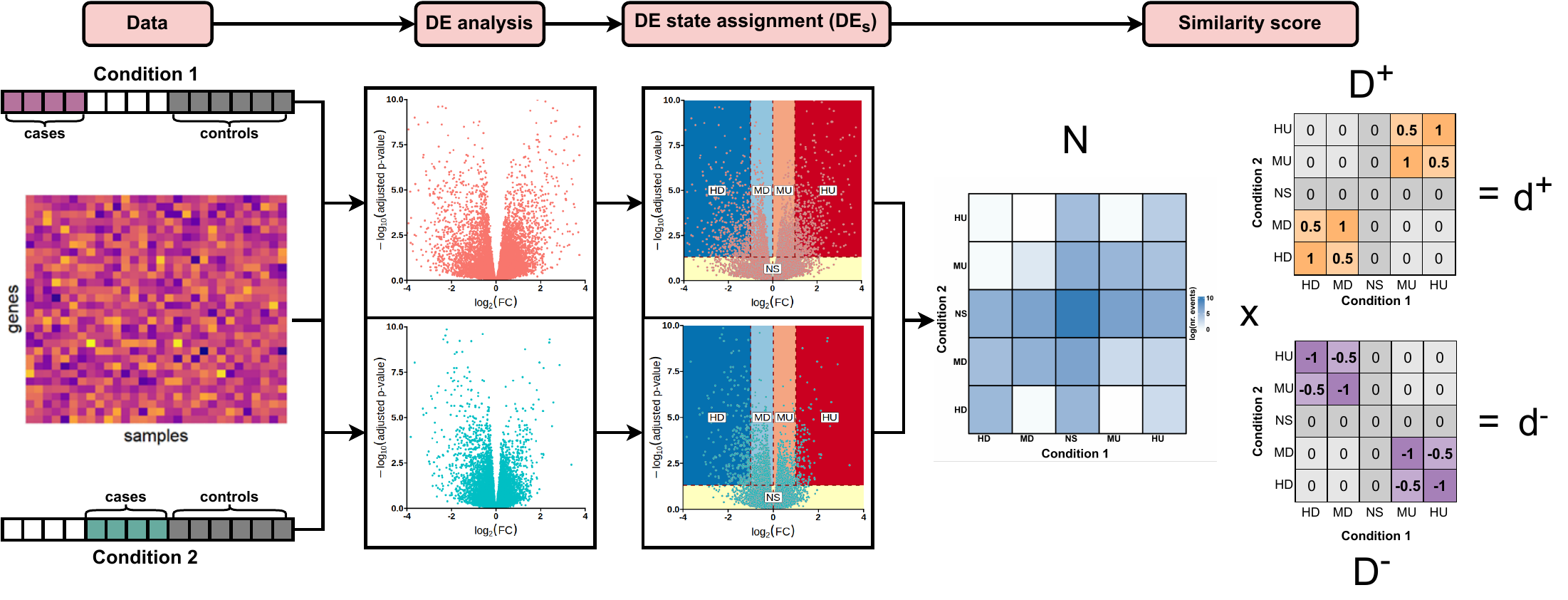}}
\caption{Flowchart of the CIBRA similarity score calculation. Expression data and group definitions with shared controls are used to perform differential expression analysis (DE) similar to the impact score calculation. Using the generated $log_2$ fold changes (FC) and adjusted p-values, DE states ($DE_s$) are defined for each gene given the corresponding p-value and fold change for both conditions.  To calculate the similarity between the two conditions, a contingency table ($N$) is generated between the $DE_s$ of the two conditions. The frequency of the combination of $DE_s$ is shown by a heatmap. A similarity ($D^+$) and anti-similarity ($D^-$) weight matrix is multiplied with the contingency table ($N$) to calculate the directional similarity scores $d^+$ and $d^-$. A permutation test is performed to estimate the significance of the $d^+$ and $d^-$ scores.}
\label{fig:flowchart_similarityscore}
\end{figure*}

 To assess the similarity in system-wide expression impact between two alterations, e.g., two different variants within the same gene, we derived a similarity score. The similarity score uses the adjusted p-values (p) and $log_2$ fold changes (FC) generated from DE analysis to estimate the distance between two alterations within this space.  For DE analysis, shared controls are needed as the similarity score assumes a shared reference point as shown in the flow chart depicted in Fig. \ref{fig:flowchart_similarityscore}. To calculate the CIBRA similarity score, first, the p-value/fold change space was divided into 5 regions: highly upregulated (HU), moderately upregulated (MU), highly downregulated (HD), moderately downregulated (MD) and not significant (NS)  (Fig. \ref{fig:flowchart_similarityscore}). Genes within this space are assigned DE states ($DE_s$) using the boundaries in equation \ref{eq4}. The equation assigns a score based on the coordinates of the gene within the fold change (FC) and adjusted p-value (p) space for the given condition. Following this step, a vector of $DE_s$ is generated for each condition. The boundaries that divide the regions are common demarcations for volcano plots in RNA-Seq data analysis to identify substantially changed genes.

\begin{equation}
 DE_s(p, FC) = \begin{cases} 
      HU & \text{if $FC \geq 1$ and $p\leq 0.05$} \\
      MU & \text{else if $0 < FC < 1$ and $p\leq 0.05$} \\
      NS & \text{else if $p > 0$} \\
      MD & \text{else if $0 > FC > -1$ and $p \leq 0.05$} \\
      HD & \text{else if $FC \leq -1$ and $p \leq 0.05$}
   \end{cases} \label{eq4}
\end{equation}

To calculate the similarity between the two conditions, first, the frequency of  $DE_s$ combinations between the two conditions is calculated. This contingency table is a 5x5 matrix termed $N$. A weight matrix representing the similarity ($D^+$) and anti-similarity ($D^-$) relationships between the $DE_s$ is multiplied with the contingency table ($N$) to calculate the directional similarity scores $d^+$ and $d^-$ as described with equations \ref{eq5} and \ref{eq6} and shown in Fig. \ref{fig:flowchart_similarityscore}.

    \begin{equation}
        d^+ = \sum_{i=1}^s\sum_{j=1}^sD_{ij}^+N_{ij} \label{eq5}
    \end{equation}
    \begin{equation}
        d^- = \sum_{i=1}^s\sum_{j=1}^sD_{ij}^-N_{ij} \label{eq6}
    \end{equation}

To assess the significance of the similarity between the two conditions, a permutation test was performed. A total of 10000 condition permutations were calculated to generate a positive and negative directional similarity score distribution. Random gene definitions from genome-wide screens were taken as conditions for permutations with shared controls. To assess the correlation between the $DE_s$ of the two conditions, the Spearman correlation measure was calculated to assess the correlation between the two $DE_s$ vectors. Given that the $DE_s$ are discrete, a Spearman correlation measure was deemed more suitable. 

\subsection{Machine Learning}
To assess if the impact of a genomic alteration could also be investigated by a machine learning model, we trained a random forest model that predicts the genomic alteration status from transcriptomics data. If the transcriptomics data contain an expression signal associated with the alteration, it should be possible to train such a machine-learning model with a reasonable performance. In this work, transcriptomics data has been used as a measure of the system changes.

\subsubsection{RNA-seq processing}
RNA-seq counts were pre-processed by removing zero variance genes and mapping ENSEMBL identifiers to HUGO gene symbols. Transcripts without HUGO gene symbol annotations were removed. The highest expressed transcript per gene was retained as the gene count. The count data were normalized using the TMM normalization method from the R package edgeR (version 3.40.1, default parameters, \cite{robinson_edger_2010}) and transformed to scaled log counts per million (logCPM) values.

\subsubsection{Random forest model}
A Random Forest classifier was built using the Classification and Regression Training R package caret (version 6.0.93, \cite{kuhn_mutation-specific_2021}). The model was trained using a 70\%/30\% train-test split with a 10x repeated 5-fold cross-validation (CV) loop on the training set for feature selection and parameter tuning. Recursive feature elimination (RFE)  was performed with a 10-300 feature range with 25 steps. The area under the precision-recall curve (PRCAUC) calculated with the R package PRROC (version 1.3.1, \cite{grau_prroc_2015}) was used as a performance metric. Parameter tuning was performed using the internal tuning step of the caret train function with a 10-value vector for the parameter 'mtry'. The final model performance was evaluated with the 30\% test set using the area under the receiver operating characteristic curve (ROCAUC) and the PRCAUC as performance measures.

\subsubsection{Permutation testing}
To assess if the performance of the ML model was significantly better than random, a permutation test was performed with 100 class label permutations. The performance of the original model was compared to the permutation performance distribution. A p-value was calculated as the fraction of permutations with a performance higher than or equal to the original model.

\subsection{Data}
\subsubsection{The Cancer Genome Atlas}
Public data from The Cancer Genome Atlas (TCGA) were gathered from the Genome Data Commons (GDC) portal for 33 cancer types \cite{grossman_toward_2016}.  Available processed single nucleotide variant calls using whole-exome sequencing data were retrieved using the R package TCGAbiolinks (version 2.25.3, \cite{colaprico_tcgabiolinks_2016}), with query parameters: data.category = ”Simple Nucleotide Variation”, data.type = ”Masked Somatic Mutation”, legacy = FALSE, access = ”open” and workflow.type = ”Aliquot Ensemble Somatic Variant Merging and Masking”. Available RNA-Seq data were retrieved with the query parameters: data.category = ”Transcriptome Profiling”, data.type = ”Gene Expression Quantification” and workflow.type = ”STAR – Counts”. Clinical data and tumor mutational burden (TMB) were retrieved from cBioPortal \cite{cerami_cbio_2012, gao_integrative_2013, weinstein_cancer_2013}. ‘Silent’ variants indicated by the variant classification provided in the Mutation Annotation Format (MAF) file were removed from the genome-wide screen analysis. 

\subsubsection{The Hartwig Medical Foundation}
From the Hartwig Medical Foundation (HMF), whole genome sequencing (WGS) data was retrieved from 610 metastatic colorectal cancer samples, 996 metastatic breast cancer samples, and 551 metastatic lung cancer samples. RNA sequencing data was available from 394 metastatic colorectal cancer samples, 332 metastatic breast cancer samples, and 127 metastatic lung cancer samples. WGS data was processed with the PURPLE-GRIDDS-LINX pipeline from HMF as previously described \cite{priestley_pan-cancer_2019} to generate SNV, SV, and SCNA calls. RNA sequencing data were analyzed with Isofox (version 1.5, \href{https://github.com/hartwigmedical/hmftools/blob/master/isofox}{Isofox GitHub}) to generate transcript counts. ‘Silent’ variants indicated by the SnpEff (version 4.3) canonical transcript summary were removed from the SNV calls in all further analyses that indicated coding variants. SV calls were only retained when they passed all filters. 

\subsection{CIBRA genome-wide screen: application to identify the impact of known and novel genomic alterations}
To evaluate the capability of CIBRA to detect known and novel genomic alterations with a system-wide impact, a gene-level genome-wide screen was performed using data from the TCGA and HMF. Cancer driver annotations were obtained from the COSMIC Cancer Gene Census database \cite{sondka_cosmic_2018}. Only Tier 1 genes were used for annotations. Multiple testing correction was performed using the Benjamini-Hochberg correction \cite{benjamini_controlling_1995}.

\textbf{HMF:}
For the HMF data consisting of breast, lung, and colorectal cancer, gene alteration definitions were constructed on four levels: SNVs, SVs, SCNAs, and any of the aforementioned alterations. For the definition, any alteration, a binary labeling was constructed where if either an SNV, SV, or SCNA occurs in the gene, the gene is given the state '1'. If none of the alterations occur within the gene, the gene is given the state '0'. The same labeling was performed for the SNV, SV, and SCNA definitions. However, the definition is only constrained to the occurrence of the given alteration within the gene. Gene annotations were retrieved using the R package AnnotationHub (version 3.6.0, \cite{morgan_annotationhub_2022}) with the query: "AH10684".

\textbf{TCGA:}
For the TCGA data on a pan-cancer level, gene alteration definitions were only constructed for coding SNVs. Variants classified as ‘silent’ were excluded from the definitions. Binary labeling was performed where genes affected by SNVs were labeled as '1' and '0' otherwise. To reduce heterogeneity within cohorts, samples with a tumor mutational burden (TMB) $>$ 10, termed 'high' TMB samples, were excluded from the analyses. Gene annotations were retrieved using the R package AnnotationHub with the query: "AH98495"

\subsection{Refining genomic alterations on mutation type and genomic location using CIBRA }
While we can define genomic alterations at the gene level, we can also zoom in and make the alterations more concrete. In this work, two ways to refine genomic alterations were assessed:

\textbf{Mutation type:}
For mutation type, the effect of SNVs classified according to their coding effects: non-coding, synonymous, missense, splice and nonsense or frameshift, SVs defined in deletions, duplications, insertions, inversions and translocations, and SCNAs in gains and losses as depicted have been assessed. The nuanced effects of mutation types have been assessed in the genes APC, TP53, KRAS, BRAF, PIK3CA, and TTN using data from microsatellite-stable CRC (HMF). 

\textbf{Genomic Location:}
For the location of alterations, 4 levels have been defined: coding regions, exons, domains, and amino acid positions. The location annotations were retrieved using the R package AnnotationHub. For domain annotations, the R packages EnsDb.Hsapiens.v75 (version 2.99.0) and EnsDb.Hsapiens.v86 (version 2.99.0) were used. A use case of the sublocation was assessed on the gene \textit{KRAS} using data from metastatic microsatellite stable CRC (HMF). 

\subsection{CIBRA similarity score: application to assess the similarity of biologically relevant alterations}
To assess the similarity in system-wide expression change between two conditions, a similarity score was calculated as described in section \ref{section_similarity}. The similarity score was calculated for two use cases using data from metastatic microsatellite-stable CRC (HMF): KRAS codon 12 compared to codon 13 variants and KRAS codon 12 compared to BRAF codon 600 variants.
For the two use cases, the shared control was wild-type for both conditions. The CIBRA impact score was calculated for each condition of the two use cases as described in section \ref{section_cibra_score}. The generated p-values and fold changes were used to calculate the CIBRA similarity score as described in section \ref{section_similarity}.

\section{Results}\label{results}

\subsection{Computational Identification of Biologically Relevant Alterations (CIBRA)}
In this work, we developed a computational workflow, CIBRA, to identify the system-wide impact of a genomic alteration. Fig. \ref{fig:full_flowchart} shows the overall workflow of CIBRA. CIBRA takes as input gene expression data and a group definition in terms of cases and controls: for example, mutated vs. WT samples for a specific genomic alteration. These inputs are used to perform differential expression (DE) analysis between the cases and controls. The p-value distribution generated from the DE analysis is subsequently decomposed by fitting a Beta-Uniform mixture model to determine the system-wide impact. From the model, the CIBRA impact score termed the significant area is calculated by taking the integral between the Beta and Uniform components of the mixture model. Hence, the CIBRA score reflects the fraction of genes with altered expression, representing the system-wide impact of the genomic alteration. Finally, the statistical significance of the impact measure is assessed with a permutation test.

\subsection{CIBRA identifies the system-wide impact of known cancer genes}
In order to assess the ability of CIBRA to identify known tumor suppressor genes and oncogenes, the impact of coding SNVs was measured by performing genome-wide screens in 33 cancer types using data from The Cancer Genome Atlas (TCGA). First, coding SNVs retrieved from processed whole exome sequencing data (WES) were grouped at the gene level. A gene is considered to be affected in a sample if a coding SNV is present within the gene. The system-wide impact was evaluated for any protein-coding gene within the genome with at least 10 cases within the cancer type tested using transcriptomics data. Of all genes affected by coding SNVs, 8.9\% (92) were identified to have a significant system-wide impact with an adjusted p-value $\leq$ 0.01.  From these genes, 54.3\% (50) are experimentally validated oncogenes and tumor suppressor genes registered in the COSMIC cancer gene census \cite{sondka_cosmic_2018}. An overview of the top 50 genes ordered on the CIBRA impact score is shown in Fig. \ref{fig:genome-screen}A. Of the 50 top scoring genes, most (80\%) are known cancer genes.  However, some genes such as General Transcription Factor IIi (\textit{GTF2I}) and Collagen Type XI Alpha 1 Chain (\textit{COL11A1}) are not (yet) registered as known tumor suppressor genes and oncogenes in the COSMIC cancer gene census database. Nevertheless, \textit{GTF2I} has gained recent attention as a characteristic cancer gene associated with spindle cell morphology in thymomas \cite{giorgetti_human_2022, petrini_specific_2014, bille_gtf2i_2022}. Similarly, mutated \textit{Col11A1} has been reported to accelerate neoplastic invasion in cutaneous squamous cell carcinomas \cite{lee_mutant_2021}. In general, these results show that CIBRA is capable of detecting the impact of known biologically relevant genes affected by SNVs.

\subsection{Structural variants have a significant system-wide impact in metastatic cancer}
\begin{figure*}
    \centering
    \captionsetup{font = footnotesize}
    \noindent\makebox[\textwidth]{\includegraphics[width=1\textwidth]{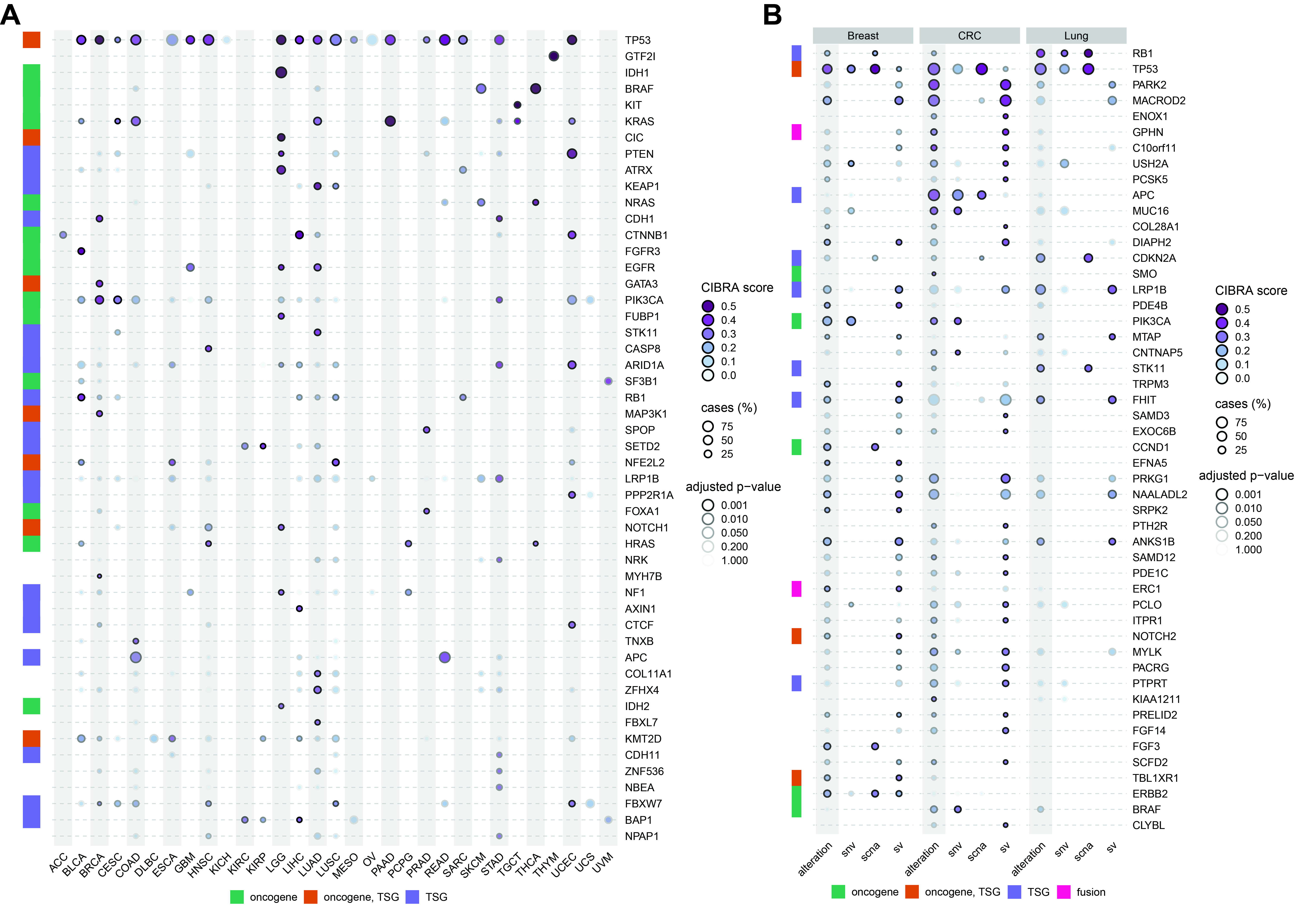}}
    \caption{Sorted gene overview of the top 50 CIBRA impact-scoring genes from genome-wide screens. \textbf{A)} An overview of CIBRA impact scores for genes affected by SNVs for primary cancer samples from 33 cancer types (TCGA). High-TMB samples ($>$ 10 TMB) were excluded to create more homogeneous cohorts. The genes were sorted by impact score. 
    \textbf{B)} An overview of CIBRA impact scores for genes affected by genomic alterations in metastatic breast, colorectal (CRC) and lung cancer (HMF).  The CIBRA score for any type of genomic alteration (highlighted in gray) and specific alterations (SNV, SCNA, and SV) are shown.
    Row annotations indicate the driver role of genes classified in tumor suppressor genes (TSG; purple), oncogenes (green), either (orange), and fusions (pink) retrieved from the COSMIC cancer gene census database. Dot color indicates the CIBRA impact score, the dot edge color indicates the significance of the CIBRA impact score assessed with a permutation test, and the size of the dot indicates the prevalence (\%) of genomic alteration with a minimum of 10 cases.}
    \label{fig:genome-screen}
\end{figure*}

Next, we assessed the biological impact of genomic alterations that have been difficult to find using frequency-based detection methods, such as SCNAs and SVs.  A genome-wide screen of genes affected by SNV, SCNA, SV, or any of the alterations mentioned before was performed on metastasized breast, lung, and colorectal cancer (CRC) using data from the Hartwig Medical Foundation (HMF). The availability of both deep-WGS data and matched RNA-seq data allowed us to assess the impact of SNVs, SCNAs, and SVs using CIBRA. From the genome-wide screen, 16.5\% (598) of genes affected by any of the three alteration types were identified to have a significant system-wide impact with an adjusted p-value $\leq$ 0.01 (Fig. \ref{fig:genome-screen} B. SVs show to often have a large effect on the system, with 30.3\% (421) of genes affected by SVs having a significant system-wide impact, especially in CRC (Fig. \ref{fig:genome-screen} B. However, from the list of alterations with a significant system-wide impact, only 12.4\% (74) are registered in the COSMIC cancer gene census. After the inclusion of SVs and SCNAs, most significant CIBRA hits are not registered as oncogenes and tumor suppressor genes (Fig. \ref{fig:genome-screen}B). Only 8.8\% (37) of the genes affected by SVs were registered. In particular, two genes located within genomic regions referred to as 'common fragile sites', Mono-ADP Ribosylhydrolase 2 (\textit{MACROD2}) and  Parkin RBR E3 Ubiquitin Protein Ligase (\textit{PRKN}, also known as \textit{PARK2}) are among the top 10 genes. Although common fragile sites are genomic regions prone to accumulate SVs under replicative stress, \textit{PRKN} shows a significant CRC-specific signal. Moreover, \textit{MACROD2} shows a significant effect in CRC and breast cancer. These results show that SVs have a significant system-wide impact in colorectal, breast, and lung cancer to a degree similar to that of known oncogenes and tumor suppressor genes. 

\subsection{Assessing gene impact by coding effect and SCNA status}
Thus far, we have explored the impact of genomic alterations at the gene level. However, CIBRA can also be used to investigate if a certain type of mutation has more impact than other alterations in the same gene. Here, we explore how CIBRA can be used to identify the most impactful mutation type by systematically testing different SNVs, classified according to their coding effect, for known tumor suppressor genes (\textit{APC} and \textit{TP53}) and oncogenes (\textit{BRAF}) in metastasized microsatellite stable CRC.

First, the impact of SNVs classified according to their coding effect was assessed. The overall system-wide impact of SNVs can be decomposed into their underlying coding effects showing stronger and weaker signals (Fig. \ref{fig:zoom_effect}A). For example, if we consider all SNVs in \textit{BRAF}, we obtain a non-significant CIBRA score (0.144, p $>$ 0.05). However, if we focus only on missense variants, a significant high impact score can be observed (CIBRA score: 0.360, p $\leq$ 0.0001). This is consistent with the expectations of the oncogene \textit{BRAF}, where missense variants, specifically V600E, have been reported to be oncogenic variants in colorectal cancer \cite{kopetz_encorafenib_2019, loupakis_kras_2009, davies_mutations_2002}. On the other hand, the frequently mutated gene \textit{TTN} shows no improvement in signal when refining the type of mutation in colorectal cancer. This aligns with the notion that due to the size of \textit{TTN} there is an accumulation of mutations associated with the tumor mutational burden without a defined effect in colorectal cancer \cite{oh_spontaneous_2020, muzny_comprehensive_2012}. A trend similar to \textit{BRAF} can be observed in \textit{TP53}, where the splice variants show the highest impact score (CIBRA score: 0.334, p $\leq$ 0.0001). However, for tumor suppressor genes such as  \textit{APC} and \textit{TP53} another refinement layer is needed. If we include SCNA status along with the SNV coding effect, the CIBRA impact score increases for both \textit{APC} and \textit{TP53} (Fig. \ref{fig:zoom_effect}B). A loss in conjunction with a coding SNV shows the highest CIBRA impact score. For example, for \textit{APC}, the highest impact score was reached with the combination of nonsense or frameshift variants and copy number loss (CIBRA score: 0.378, p $<$ 0.0001), while with only nonsense or frameshift variants, the impact score was similar to the effect of only copy number loss (CIBRA score: 0.245, p $<$ 0.005). A similar effect can be observed for \textit{TP53}. This reaffirms the notion that tumor suppressor genes require two hits to have the most effect and shows that CIBRA can detect this property \cite{ashley_two_1969, datta_tumor_2020, berger_continuum_2011}. Hence, refining on coding effect enables us to observe which type of alterations have the most impact within a gene.

\subsection{Refining gene impact by zooming in on genomic location}
\begin{figure*}
        \centering
        \captionsetup{font = footnotesize}
        \noindent\makebox[\textwidth]{\includegraphics[width=0.8\textwidth]{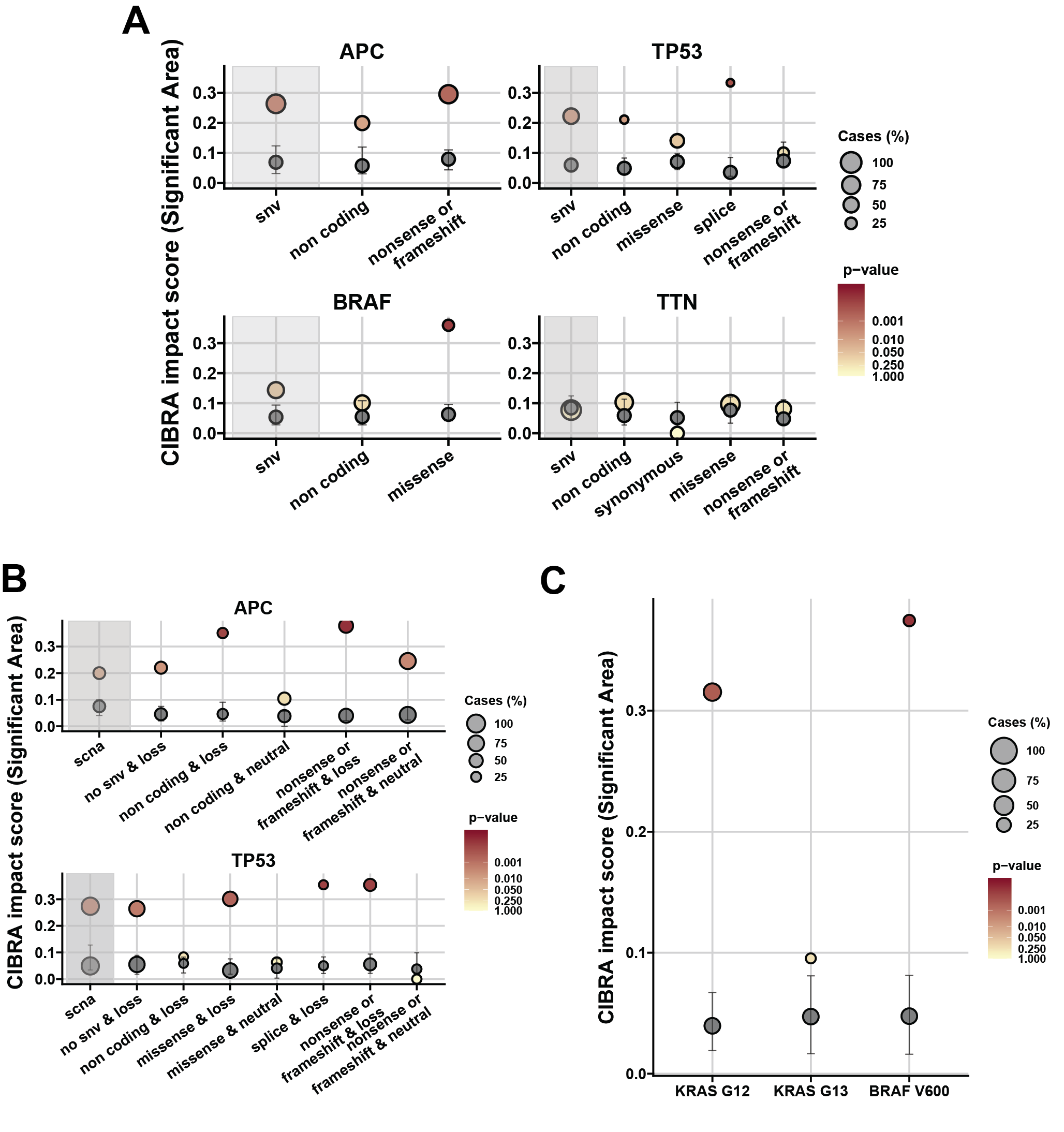}}
        \caption{Refining genomic alterations by mutation type and genomic location for the identification of impactful alterations using CIBRA.  \textbf{A)} CIBRA impact score for SNVs dissected in their underlying coding effects: non-coding, synonymous, missense, splice, and nonsense or frameshift for 4 frequently mutated genes in metastatic colorectal cancer: \textit{APC}, \textit{TP53}, \textit{BRAF}, and \textit{TTN} using data from metastatic microsatellite stable CRC.  Two of the genes, \textit{APC} and \textit{TP53}, are known tumor suppressor genes. \textit{BRAF} is a known oncogene, and \textit{TTN} is a recurrently mutated gene without a known cancer-associated function in CRC.  \textbf{B)} CIBRA impact score for SCNAs combined with SNV coding effects for the TSGs \textit{APC} and \textit{TP53} in metastatic microsatellite stable CRC. \textbf{C)} CIBRA impact score visualization of \textit{KRAS} codon 12 (G12), 13 (G13) and \textit{ BRAF} codon 600 (V600) mutations. \textbf{A, B \& C)} Shaded gray area: the impact score of the overall effect of SNVs or SCNAs. Gray dot: median permutation CIBRA score, with the error bars representing the IQR. Dot color: significance of the CIBRA score assessed by a permutation test. Dot size: prevalence of the alteration (\%).}
        \label{fig:zoom_effect}
    \end{figure*}

The most impactful mutation within a cancer gene could be identified by refining on mutation type. However, the genomic location of an alteration can also influence the impact. To exemplify this, we explored the impact of mutations within subregions in \textit{KRAS} and \textit{BRAF}. 
If we focus on \textit{KRAS} and dissect the coding variants on their genomic location, only codon 12 (G12) variants showed a significant system-wide impact in metastatic microsatellite stable CRC (CIBRA score: 0.32, p $<$ 0.001; Fig. \ref{fig:zoom_effect}C). This is in line with the distinct clinical behavior between codon 12 and 13 \cite{jones_specific_2017, yoon_kras_2014}. Moreover, codon 12 variants show an impact to a similar extent as \textit{BRAF} codon 600 variants (CIBRA score: 0.37, p $<$ 0.001; Fig. \ref{fig:zoom_effect}C). 

\subsection{Distinct expressional similarity pattern between \textit{KRAS} and \textit{BRAF} coding variants in metastatic colorectal cancer}
To assess if two alterations have the same  gene expression patterns, CIBRA can derive a similarity score between the distinct genomic alterations.
The CIBRA similarity score compares two genomic alterations in contrast to the same control condition and quantifies the extent of similarity between the two conditions. For the CIBRA similarity score, group definitions with shared controls are used to first perform differential expression analysis. Second, the generated $log_2$ fold changes and adjusted p-values are used to assign genes per condition to differential expression states. The differential expression states are five directional states that range from highly downregulated to highly upregulated (Fig. \ref{fig:full_flowchart_bottom}). By capturing the DE genes in these five categories, only large changes will be considered in the similarity score. Last, a similarity and anti-similarity score is calculated from the differential expression states (Fig. \ref{fig:full_flowchart_bottom}). 

\begin{figure*}
    \captionsetup{font = footnotesize}
    \centering
    \noindent\makebox[\textwidth]{\includegraphics[width=\textwidth]{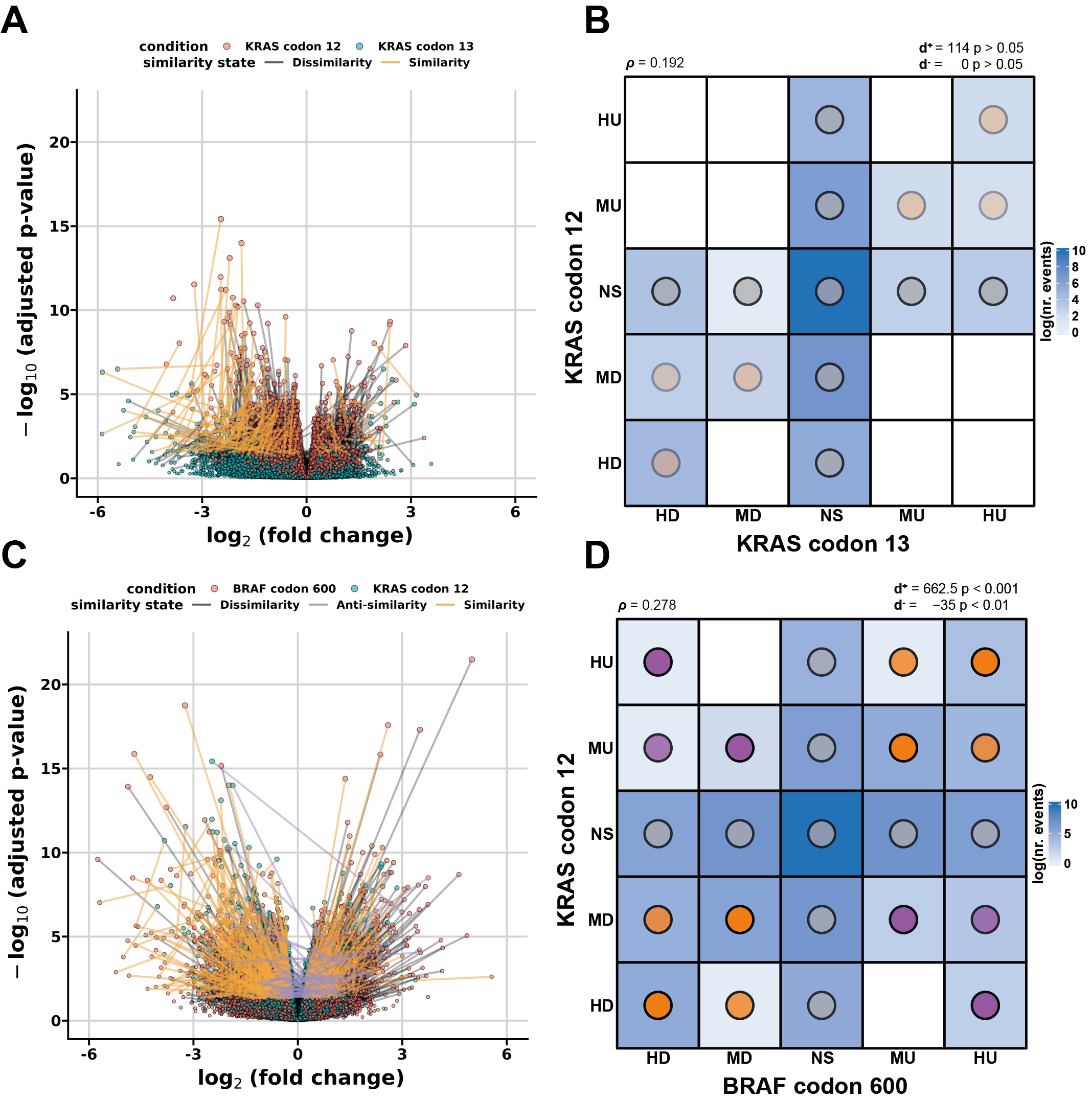}}
    \caption{Biological similarity assessment between two genomic alterations (conditions) in metastatic microsatellite stable colorectal cancer (HMF). \textbf{A}) Linked volcano plot showing the $log_2$ fold change (FC) and the adjusted p-value of genes differentially expressed between \textit{KRAS} G12 variants (red) and WT and \textit{KRAS} G13 variants (blue) and WT. The similarity (orange), anti-similarity (purple), and dissimilarity (gray) states are shown by the colored links between the corresponding genes. \textbf{B}) Categorical DE similarity pattern between \textit{KRAS} G12 and G13 variants compared to \textit{KRAS} WT. The color indicates the log-transformed number of events (genes) corresponding to the DE state combinations. The dot color indicates the type of similarity present in the pattern, and the intensity is proportional to the p-value of the similarity score with an upper bound of 0.001. The directional similarity has been reported as similarity and anti-similarity with the corresponding p-value assessed with a permutation test. \textbf{C}) Linked volcano plot showing the $log_2$ fold change (FC) and adjusted p-value of genes differentially expressed between \textit{BRAF} V600 variants and \textit{KRAS} and \textit{BRAF} WT (red) and \textit{KRAS} G12 variants and WT (blue). \textbf{D)} Categorical DE similarity pattern between \textit{KRAS} G12 and \textit{BRAF} V600 variants compared to \textit{KRAS} and \textit{BRAF} WT samples.}
    \label{fig:genomic_similarity}
\end{figure*}

To illustrate the similarity score, variants within the same gene and variants between different genes are compared. For the within-gene comparison, \textit{KRAS} codon 12 and codon 13 variants are considered using data from microsatellite stable metastatic CRC. No significant similarity was observed between codons 12 and 13 ($d^+$ = 114, p $>$ 0.05; Fig. \ref{fig:genomic_similarity}A \& B). This difference can be mainly attributed to the lack of DE genes for codon 13 that are observed for codon 12. This highlights that codon 12 variants have a stronger distinct effect on the system. However, there are also a few significant DE genes for codon 13 that are also captured by codon 12 as shown by the light diagonal line in Fig. \ref{fig:genomic_similarity}B.
On the other hand, if we focus more closely on \textit{KRAS} G12 and \textit{BRAF} V600 variants, two genes in the same pathway with a significant system-wide impact, a distinct significant pattern of similarity ($d^+$ = 662.5, p $<$ 0.001) and anti-similarity ($d^-$ = -35, p $<$ 0.01)  can be observed (Fig. \ref{fig:genomic_similarity}C \& D).  This highlights that although both genes are part of the EGFR pathway, they show a different effect on the system in metastatic microsatellite stable CRC. This is in line with the different clinical behaviors reported between \textit{KRAS} and \textit{BRAF} variants in CRC \cite{reischmann_brafv600e_2020, morkel_similar_2015}.  

\subsection{CIBRA versus Machine learning (RF)}
\begin{figure*}
    \centering
    \captionsetup{font = footnotesize}
    \includegraphics[width=0.8\linewidth]{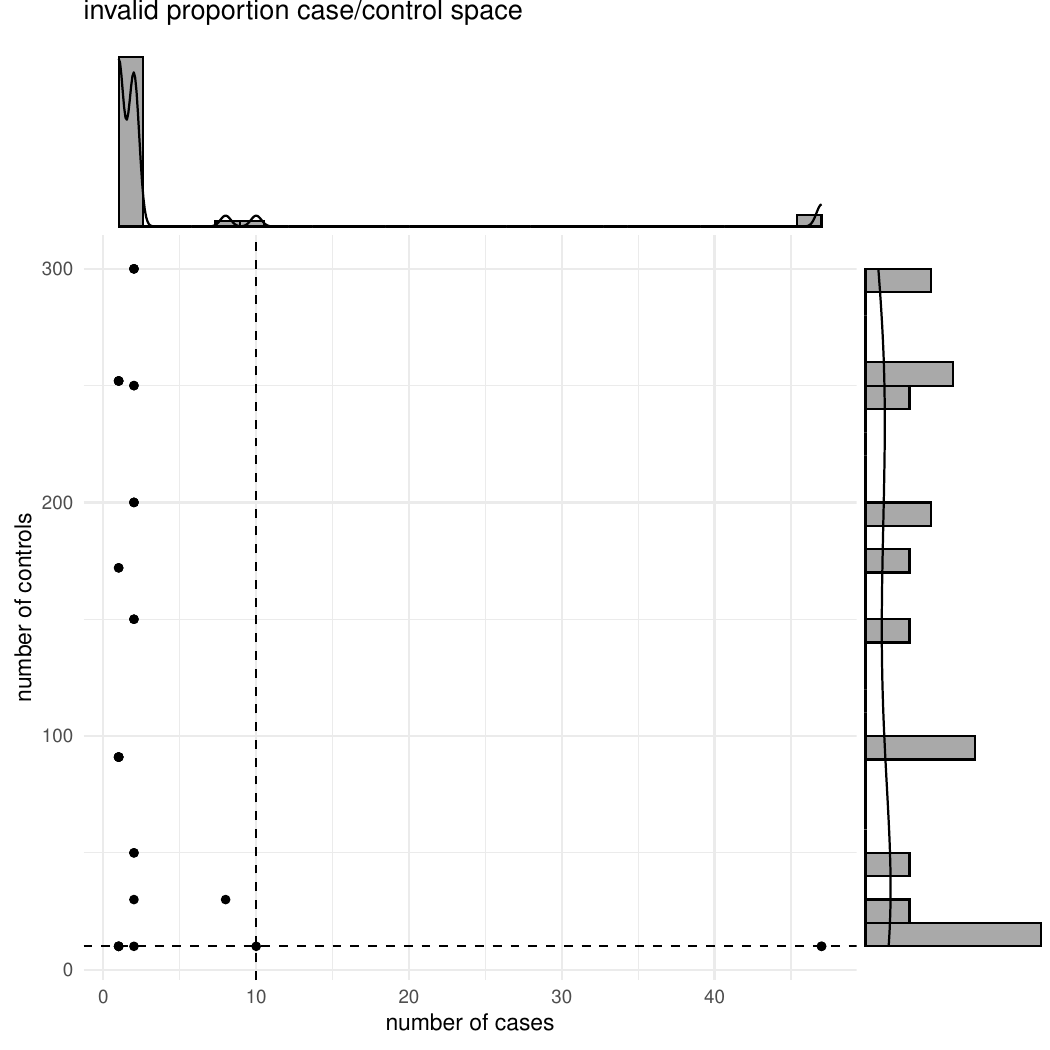}
    \caption{Number of cases plotted against the number of controls that resulted in an invalid proportion below the threshold $\tau$ assessed with 1000 sample permutations. The threshold $\tau$ used in this manuscript is 0.1. Sample permutations with fewer than 10 cases resulted in invalid signal measures. The number of cases between 3 and 8 could not be assessed with DESeq2, as they resulted in errors.}
    \label{fig:sub_4}
\end{figure*}

Last, an alternative method was evaluated to determine the impact of genomic alterations. We hypothesized that if a mutation has an impact on the expression levels of the biological system, it should be possible to predict the mutation status of a sample from the genome-wide expression profile. As such, a machine learning model was trained to predict the mutation status of five known cancer genes given transcriptomics data. The performance of such a machine learning model was compared with the CIBRA impact score for five known cancer genes in metastatic CRC: \textit{TP53}, \textit{APC}, \textit{KRAS}, \textit{BRAF}, and \textit{PIK3CA}, as well as a gene often mutated due to its size, \textit{TTN}, with no reported impact. The model performance to predict the mutation status of the six assessed genes is shown in Table \ref{table:rf_cibra}. Various definitions of mutations have been evaluated, including the combination of SNV and SCNA in the tumor suppressor genes \textit{TP53} and \textit{APC}. The mutation status of the cancer genes is indeed predictable from their expression profiles as shown by the receiver operating characteristic area under the curve (ROC AUC) scores (Table \ref{table:rf_cibra}). However, not all mutation definitions could be assessed using the machine learning approach, as the machine learning model needed at least 50 cases and controls to be reasonably trained and validated. In contrast, the CIBRA impact score can be calculated for all genomic alterations in table \ref{table:rf_cibra}, and could calculate the impact score with a minimum of 10 cases and controls (Fig. \ref{fig:sub_4}). Moreover, CIBRA was able to identify the system-wide impact of all cancer genes evaluated and confirms that TTN does not have a system-wide impact in CRC, as also observed with the machine learning model. Overall, these results show that both machine learning models and the CIBRA impact score can be used as impact indicators for genomic alterations, while the CIBRA impact score is preferred when a small number of samples are available for a specific alteration.

\section{Discussion}\label{sec12}
Identifying alterations with a system-wide impact is a challenging task. CIBRA integrates two different omics data types to determine the system-wide impact of genomic alterations. By integrating multi-omics data, CIBRA was able to identify the majority of known tumor suppressor genes and oncogenes from a pan-cancer genome-wide screen of primary cancers.  Notably, when we applied CIBRA on a genome-wide screen of genes affected by SNVs, SCNAs, and SVs using data from metastatic colorectal, breast, and lung cancer, many genes affected by SVs showed a significant system-wide impact to a similar extent to known cancer genes such as  \textit{TP53} and \textit{APC}. This finding suggests that the impact of structural variants has been largely underestimated. Furthermore, CIBRA could systematically refine genomic alterations on genomic sublocations and mutation types to narrow impactful mutations. Finally, CIBRA could facilitate the evaluation of similarities between genomic alterations in terms of their impact. The systematic identification of the most impactful type of mutation and gene subregion can facilitate the prioritization and design of focused experimental assays to validate the observed system-wide impact and allow for the reassessment of their clinical implications. 

From the pan-cancer genome-wide screen of primary cancers, many tumor suppressor genes and oncogenes showed a significant system-wide change in the transcriptome.  We hypothesized that biologically relevant alterations elicit large changes throughout the system. As hallmarks of cancer are central components of the cell and the tumor, disruption of these processes can induce large trickle-down effects through the system, explaining the many known cancer genes found with a large system-wide effect \cite{hanahan_hallmarks_2022, hanahan_hallmarks_2011}. Although cancer genes are often key components of hallmarks of cancer, genomic alterations within these genes do not necessarily necessitate a large change in the system \cite{sanchez-vega_oncogenic_2018}. CIBRA was able to detect these trickle-down effects from cancer genes and verified our hypothesis that cancer genes can elicit such large effects through the system. Notably, not all of the highest-scoring genes are registered tumor suppressor genes and oncogenes, such as \textit{GTF2I} and \textit{COL11A1}. However, even though they are not registered in the cosmic cancer gene census, recent studies have shown the tumorigenic potential of \textit{GTF2I} in thymoma and \textit{COL11A1} in squamous cell carcinomas \cite{giorgetti_human_2022, petrini_specific_2014, bille_gtf2i_2022, liu_thymic_2022, lee_mutant_2021}. On the other hand, from our metastatic breast, colorectal, and lung cancer screens, SVs emerged as having a significant system-wide impact on cancer. Notably, genes within common fragile sites, such as \textit{PRKN} and \textit{MACROD2}, which are among the top 10 scoring genes, showed an exorbitant impact in metastasized CRC, highlighting their potential relevance in cancer. \textit{MACROD2} has been reported to promote chromosomal instability, positioning it as a potential tumor suppressor gene \cite{sakthianandeswaren_macrod2_2018}.  Variants within \textit{PRKN} have been reported to result in mitotic instability, contributing to oncogenesis \cite{veeriah_somatic_2010}. Our findings suggest that the impact of SVs in cancer has thus far been underreported, and more attention is needed for SVs in the field of cancer biology and precision oncology.

Although the system-wide impact can be the result of changes at different levels of the system, from cells to tissue and its microenvironment. There are also some limitations to the CIBRA methodology. First, an absence of impact is not an indication of no importance, as the variant may play a relevant role in another setting, such as early in cancer progression. When assessing the impact of the variant in the metastasized setting, the impact of the variant may be overlooked. Moreover, the homogeneity of the set of samples influences the power of the CIBRA impact and similarity scores. The more homogeneous the set of samples, the better the signal-to-noise ratio becomes. Finally, the pattern of similarity depends on the setting of the samples used. To gain a full understanding of the similarity pattern, different contexts have to be assessed, such as different cancer types or stages. Overall, both the systematic refinement of mutation type and gene subregion and the similarity score are powerful tools to help improve our understanding of cancer biology. 

In this work, we have shown that the impact of genomic alterations can be determined in two ways by integrating multi-omics data: a statistical method that determines the system-wide impact of genomic alterations, i.e., CIBRA, and a machine learning model that identifies characteristic changes in the system associated with the genomic alteration. Machine learning methods such as a random forest model have the strength that they can identify non-linear relationships that do not necessarily consist of system-wide effects, allowing them to identify a wider range of cancer gene effects. However, these methods require sufficient data to train and validate a model, limiting their scope of use. In contrast, CIBRA is able to determine the system-wide impact of any genomic alteration with at least 10 cases and controls, allowing us to explore a wider and deeper space, given the assumption that the genomic alteration has its impact in a linear system-wide effect.  

Systematically refining the definition of alterations on mutation type and genomic location can be helpful in two aspects: first, to prioritize and design focused experiments for validating causative relationships, and second, to select potential biomarkers for clinical testing and validation. Due to the limited size of patient cohorts, clinical data is often insufficient to fully explore the potential biomarker space, limiting the identification or refinement of potential biomarkers \cite{ren_pitfalls_2020}. CIBRA enables the prioritization of biomarkers by pre-selecting and refining their definition. In this way, noisy definitions can be cleared to allow for more effective testing and validation of clinical markers with limited clinical data. In this scenario, clinical data are used more effectively to test and validate potential biomarkers than for discovery. Therefore, a preliminary screening and refinement of genes affected by alterations, as demonstrated in this manuscript, can provide a valuable first step to direct further exploration and validation for the identification of biomarkers for personalized care.

In conclusion, CIBRA is a versatile method for identifying genomic alterations with a system-wide impact on tumor biology and has the potential to help in our understanding of disease-associated genomic alterations.

\backmatter

\bmhead{Data availability}
All datasets analyzed in this study are publicly available. The data generated in this study are available within this article, its supplementary data files, and the code and data repositories.

\bmhead{Code availability}
Our R package CIBRA is available at \url{https://github.com/AIT4LIFE-UU/CIBRA}. 

\bmhead{Acknowledgments}
This publication and the underlying study have been made possible partly through data requested (DR-072) and made available by the Hartwig Medical Foundation (HMF) and the Center of Personalized Cancer Treatment (CPCT). The results shown here are in part based on data generated by the TCGA Research Network (https://www.cancer.gov/tcga). We thank Libio Goncalves Braz from the department of Information and Computing Sciences from Utrecht University for testing the usability of the package on multiple systems.

\bmhead{Author Contributions} S.L. has contributed to the conceptualization, formal analysis, validation, investigation, visualization, methodology, writing of the original draft, and software development. C.B. has contributed to the formal analysis, investigation, methodology, and review of the writing. G.A.M. and J.H. have contributed to the conceptualization, supervision, review of the writing, and funding acquisition. R.J.A.F. has contributed to the conceptualization, supervision, funding acquisition, writing of the original draft, and review of the writing. S.A. has contributed to the conceptualization, supervision, funding acquisition, writing of the original draft, review of the writing, and methodology.

\bmhead{Competing interests}
R.J.A.F. reports grants and non-financial support from Personal Genome Diagnostics, non-financial support from Delfi Diagnostics, grants from MERCK BV, grants and non-financial support from Cergentis BV, outside the submitted work; In addition, R.J.A.F. has several patents pending.
S.A. reports grants and non-financial support from Cergentis BV and a patent pending, outside the submitted work. S.L. reports non-financial support from Cergentis BV and a patent pending, outside the submitted work. J.H. reports a patent pending, outside the submitted work. G.A.M. is co-founder and board member (CSO) of CRCbioscreen BV, CSO of Health-RI (Dutch National Health Data Infrastructure for Research \& innovation), and member of the supervisory board of IKNL (Netherlands Comprehensive Cancer Organisation). G.A.M. non-financial support from Exact Sciences, non-financial support from Sysmex, non-financial support from Sentinel CH. SpA, non-financial support from Personal Genome Diagnostics (PGDX), non-financial support from DELFI, other from Hartwig Medical Foundation, grants from CZ (OWM Centrale Zorgverzekeraars groep Zorgverzekeraar u.a), other from Royal Philips, other from GlaxoSmithKline, other from Keosys SARL, other from Open Clinica LLC, other from Roche Diagnostics Nederland BV, other from The Hyve BV, other from Open Text, other from SURFSara BV, other from Vancis BV, other from CSC Computer Sciences BV, outside the submitted work; In addition, G.A.M. has several patents pending. The other authors declare no potential conflicts of interest.

\bibliography{impact_lib}

\fancypagestyle{plain}{%
  \fancyhf{}
  \fancyfoot[C]{\iffloatpage{}{\thepage}}
  \renewcommand{\headrulewidth}{0pt}}
  
\newpage
\beginsupplement 
  \pagestyle{plain}

\section*{Supplemental material}
\begin{table*}[h!]
   \captionsetup{font = footnotesize}
   \caption{Comparing the CIBRA score and machine learning models to identify the impact of genomic alterations using transcriptomics data. Both the CIBRA impact score and a random forest (RF) classification model performance score were used to assess the impact of coding SNVs, SCNAs, the combination (SNV + SCNA), or any alteration in \textit{TP53}, \textit{APC}, \textit{KRAS}, \textit{BRAF}, \textit{PIK3CA}, and \textit{TTN}. The impact is measured with the significant area for CIBRA and the receiver operating characteristic area under the curve (ROC AUC) for the RF model. Significance is determined through a permutation test for both methods.}    
   \centering
    \noindent
   \begin{adjustbox}{center}
    \begin{tabularx}{1.2\textwidth}{@{} >{\centering\arraybackslash}m{3.2em}>{\centering\arraybackslash}m{6.2em}>{\centering\arraybackslash}m{5em}>{\centering\arraybackslash}m{8 em}>{\centering\arraybackslash}m{3.8em}>{\centering\arraybackslash}m{6em}>{\centering\arraybackslash}m{3.8em} @{}}
       \toprule
        \multicolumn{3}{N }{\textbf{}} & \multicolumn{2}{c }{\textbf{CIBRA}} & \multicolumn{2}{c }{\textbf{Machine Learning (RF)}} \\
        \cmidrule(lr){4-5}
        \cmidrule(ll){6-7}
        \textbf{Gene} & \textbf{Alteration} & \textbf{Cases (samples)} & \textbf{Significant area (score)} & \textbf{p-value} & \textbf{ROCAUC (score)} & \textbf{p-value} \\
        \cmidrule(lr){1-3}
        \cmidrule(lr){4-5}
        \cmidrule(ll){6-7}
        \emph{TP53} & only SNV & 35 & 0.094 & 0.38 & - & - \\
        ~ & only SCNA & 133 & 0.298 & 0.007 & 0.92 & $\leq$ 0.01 \\
        ~ & SNV \& SCNA & 144 & 0.378 & 0.001 & 0.94 & $\leq$ 0.01 \\
        ~ & any alteration & 327 & 0.327 & 0.004 & 0.95 & $\leq$ 0.01 \\
       \cmidrule(lr){1-3}
       \cmidrule(lr){4-5}
       \cmidrule(ll){6-7}
       \emph{APC} & only SNV & 173 & 0.306 & 0.006 & 0.86 & $\leq$ 0.01 \\ 
       ~ & only SCNA & 48 & 0.329 & 0.004 & - & - \\ 
       ~ & SNV \& SCNA & 67 & 0.383 & 0.001 & 0.88 & $\leq$ 0.01 \\ 
        ~ & any alteration & 303 & 0.357 & 0.003 & 0.80 & $\leq$ 0.01 \\ 
        \cmidrule(lr){1-3}
        \cmidrule(lr){4-5}
        \cmidrule(ll){6-7}
        \emph{KRAS} & only SNV & 163 & 0.210 & 0.05 & 0.84 & $\leq$ 0.01 \\ 
        ~ & any alteration & 165 & 0.212 & 0.04 & 0.81 & $\leq$ 0.01 \\ 
        \cmidrule(lr){1-3}
        \cmidrule(lr){4-5}
        \cmidrule(ll){6-7}
        \emph{BRAF} & only SNV & 51 & 0.280 & 0.01 & 0.74 & $\leq$ 0.01 \\ 
        ~ & any alteration & 61 & 0.214 & 0.04 & 0.75 & $\leq$ 0.01 \\ 
        \cmidrule(lr){1-3}
        \cmidrule(lr){4-5}
        \cmidrule(ll){6-7}
        \emph{PIK3CA} & only SNV & 60 & 0.325 & 0.004 & 0.61 & 0.16 \\ 
        ~ & any alteration & 61 & 0.319 & 0.004 & 0.66 & 0.01 \\
        \cmidrule(lr){1-3}
        \cmidrule(lr){4-5}
        \cmidrule(ll){6-7}
        \emph{TTN} & only SNV & 210 & 0 & 1 & 0.52 & 0.38 \\ 
        ~ & any alteration & 219 & 0 & 1 & 0.4 & 0.93 \\
        \bottomrule
   \end{tabularx}

\end{adjustbox}
\label{table:rf_cibra}
\end{table*}

\end{document}